\documentclass[conference]{IEEEtran}
\IEEEoverridecommandlockouts

\usepackage{xcolor}

\usepackage{cite}
\usepackage{amsmath,amssymb,amsfonts,bm}
\usepackage{algorithmic}
\usepackage{nicefrac}
\usepackage{lipsum}
\usepackage{graphicx}
\ifCLASSOPTIONcompsoc
    \usepackage[caption=false, font=normalsize, labelfont=sf, textfont=sf]{subfig}
\else
\usepackage[caption=false, font=footnotesize]{subfig}
\fi
\usepackage{textcomp}
\usepackage{xcolor}
\DeclareUnicodeCharacter{0308}{}

\def\BibTeX{{\rm B\kern-.05em{\sc i\kern-.025em b}\kern-.08em
    T\kern-.1667em\lower.7ex\hbox{E}\kern-.125emX}}
\begin{document}

\title{Communication-avoiding micro-architecture to compute Xcorr scores for peptide identification\\
}

\author{\IEEEauthorblockN{Sumesh Kumar, Fahad Saeed}
\IEEEauthorblockA{\textit{Knight Foundation School of Computing and Information Sciences} \\
\textit{Florida International University (FIU)}\\
Miami, FL USA 33199\\
\{sumesh.kumar, fsaeed\}@fiu.edu}
}

\IEEEoverridecommandlockouts
\IEEEpubid{\makebox[\columnwidth]{978-1-5386-5541-2/18/\$31.00~\copyright2018 IEEE \hfill} \hspace{\columnsep}\makebox[\columnwidth]{ }}

\maketitle

\begin{abstract}
Database algorithms play a crucial part in systems biology studies by identifying proteins from mass spectrometry data.  Many of these database search algorithms incur huge computational costs by computing similarity scores for each pair of sparse experimental spectrum and candidate theoretical spectrum vectors. Modern MS instrumentation techniques which are capable of generating high-resolution spectrometry data require comparison against an enormous search space, further emphasizing the need of efficient accelerators. Recent research has shown that the overall cost of scoring, and deducing peptides is dominated by the communication costs between different hierarchies of memory and processing units. However, these communication costs are seldom considered in accelerator-based architectures leading to inefficient DRAM accesses, and poor data-utilization due to irregular memory access patterns. In this paper, we propose a novel communication-avoiding micro-architecture to compute cross-correlation based similarity score by utilizing efficient local cache, and peptide pre-fetching to minimize DRAM accesses, and a custom-designed peptide broadcast bus to allow input reuse. An efficient bus arbitration scheme was designed, and implemented to minimize synchronization cost and exploit parallelism of processing elements. Our simulation results show that the proposed micro-architecture performs on average 24x better than a CPU implementation running on a 3.6 GHz Intel i7-4970 processor with 16GB memory.  

\end{abstract}


\section{Introduction}
Mass-spectrometry based analysis has been the preferred method for identification of proteins from complex biological samples \cite{aebersold2003mass}. The last two decades have seen tremendous developments in data acquisition and analysis techniques which have enabled many powerful proteomic applications. Database search algorithms such as SEQUEST\cite{eng2008fast}, X!Tandem\cite{craig2004tandem}, and MSFragger \cite{kong2017msfragger} can now search high resolution mass-spectrometry data against an ever increasing protein database to produce high quality matches. This has drastically increased the compute load for existing implementations of the database search algorithms. In this regard, several studies have used parallelization strategies using high-performing compute clusters\cite{li2019sw,wang2010efficient,sadygov2002code}, GPUs\cite{milloy2012tempest,li2014accelerating,baumgardner2011fast}, and FPGAs\cite{coca2010parallel,casasopra2016parallel,yang2019complete,qiu2017fpga} to speed up the computation process. However, a recent study suggests that the major bottleneck in mass-spectrometry based analysis is the cost of communication i.e cost of moving input and output data between different hierarchies of a system\cite{saeed2020communication}. Thus, even though CPUs are operating at a much higher frequency, their performance gain for proteomics studies relies on efficiently utilizing system cache or some other input reuse technique~\cite{asano2009performance} to minimize the number of DRAM accesses. Consequently, the implementation of Crux~\cite{mcilwain2014crux}, state-of-the-art software for computing cross-correlation (Xcorr) scores, utilizes processor registers to store peptide fragment ions to allow peptide reuse. While this allows one-side data reuse, the cost of accessing experimental spectra from main memory is not minimized as generally CPU registers are not large enough to hold the entire experimental spectrum. On the other hand custom architectures using FPGAs can achieve better performance for memory bound applications by utilizing the abundant on-chip RAM resources and custom-designed communication minimizing pipelines to allow experimental spectrum reuse\cite{nurvitadhi2016accelerating}.\par 

In this paper, we propose a communication-avoiding micro-architecture to accelerate the Xcorr score computation which achieves two-side data reuse by utilizing the on-chip RAM to cache an entire experimental spectrum and a peptide broadcast bus to decrease the number of DRAM accesses. Our experiments show that these optimizations result in 600x reduction in the average number of DRAM accesses compared with a no-caching approach and 24x times speed-up over Crux. The main contributions of this paper are as follows: 
\begin{enumerate}
    \item We implemented a block RAM based cache of size 2kB to store experimental spectra and minimize redundant DRAM accesses.
    \item We pre-sorted the peptide database which allows the use of binary search to search candidate peptides. The search operation needs to be performed only once per spectrum as next peptide can be pre-fetched, hence achieving input locality. 
    \item To allow input reuse, we designed a peptide broadcast bus to make it accessible to all the processing elements.
    \item  We implemented a first-come first-serve (FCFS) based bus arbitration scheme to minimize the synchronization time of processing elements sharing the system bus.

\end{enumerate}
\IEEEpubidadjcol

The rest of the paper is organized as follows. Section II explains the background of Xcorr computation problem and related work. Section III describes the proposed architecture. Section IV presents the experimental results. Section V concludes the paper.

\section{Background}

\subsection{Xcorr theoretical formulation}
The Xcorr score between a theoretical spectrum vector $\bm{X}$ and an experimental spectrum vector $\bm{Y}$ of length $n$ is defined in \cite{eng1994approach} as,

\begin{equation}
 X_{corr} = \sum_{i=0}^{n-1}X[i] Y[i] -   \frac{1}{151}
            \sum_{i=0}^{n-1}\sum_{\tau = -75}^{\tau = 75} X[i] Y[i-\tau] 
\end{equation}

where $\tau$ is the amount by which vector is being serially shifted. However, SEQUEST implementation performs an optimization by pre-processing the experimental spectrum to perform dot product only once as described in\cite{eng2008fast} and summarized below,
%

 
\begin{equation}
 \bm{Y_P} =  \sum_{i=0}^{n-1} \left(Y[i] -   \frac{1}{151}
            \sum_{\tau = -75}^{\tau = 75} Y[i-\tau] \right)
\end{equation}
 
 using (2) reduces the Xcorr computation to 
\begin{equation*}
X_{corr} = \sum_{i=0}^{n-1}X[i] Y_P[i]
\end{equation*}

\subsection{Related work}
A significant amount of work has been done on acceleration of peptide deduction algorithms using FPGA based architectures. Bogdan, Coca and Beynon\cite{coca2010parallel} designed a FPGA based accelerator for peptide deduction using Profound\cite{zhang2000profound} algorithm by instantiating 48 parallel search processors to achieve 950$\times$ speed-up over a single core processor. Another study accelerated the scoring process of X!Tandem\cite{qiu2017fpga} by instantiating 6 score generation modules and one fragment ion generation module to achieve 17$\times$ speed-up over a CPU only implementation. In \cite{yang2019complete}, a complete CPU-FPGA system which was based on\cite{qiu2017fpga} but included support for multiple FPGAs and achieved 10-fold speed-up for the entire search operation. To the best of our knowledge, there hasn't been any studies performed on accelerating the SEQUEST\cite{baumgardner2011fast} algorithm, which is widely used in proteomics.

\section{Proposed Architecture}
\begin{figure}[ht]
\begin{center}
\includegraphics[width=1\linewidth]{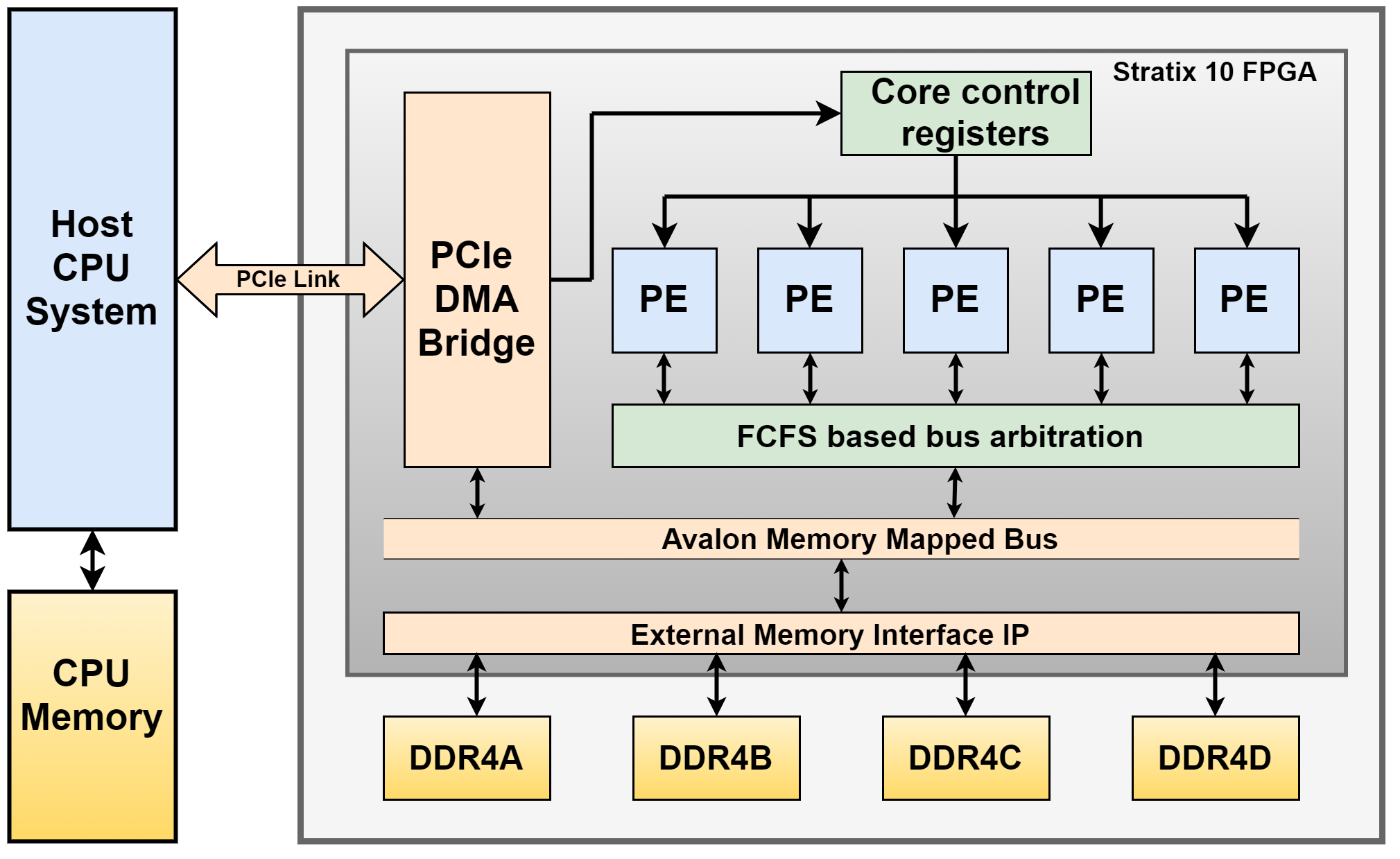}
\caption{Complete system architecture shows host CPU communicates with FPGA RAM via  PCIe DMA bridge which is connected to Intel's Avalon memory mapped bus. Core registers module contains the computation parameters and is also used for FPGA-CPU communication. To allow efficient use of the Avalon memory mapped bus, all PEs are connected to FCFS based bus arbiter which is in turn connected to Avalon memory mapped bus. }
\label{fig:system}
\end{center}
\end{figure}

The architectural setting of the heterogeneous computational system for Xcorr is shown in Fig. \ref{fig:system}. Host CPU communicates with PCIe DMA via the PCIe link to transfer the experimental spectra from host memory to FPGA memory. A set of directly accessible core control registers, hold computation parameters, and control the operation. Each step of the algorithm takes place inside the processing element (PE) i.e. reading experimental spectrum vectors one by one, searching for candidate peptides, generating theoretical spectrum, computing dot product scores, and writing the results back to main memory. The system allows deployment of multiple PEs which execute the computations in a parallel and asynchronous manner. Since all the PEs share the same memory bus, we implemented a first come first serve (FCFS) based bus arbitration scheme to achieve maximum bandwidth utilization. The detailed view of the bus arbitration scheme is described in Fig. \ref{fig:arbiter}.

\subsection{Processing Element construction}
\begin{figure}[ht]
\begin{center}
\includegraphics[width=0.85\linewidth]{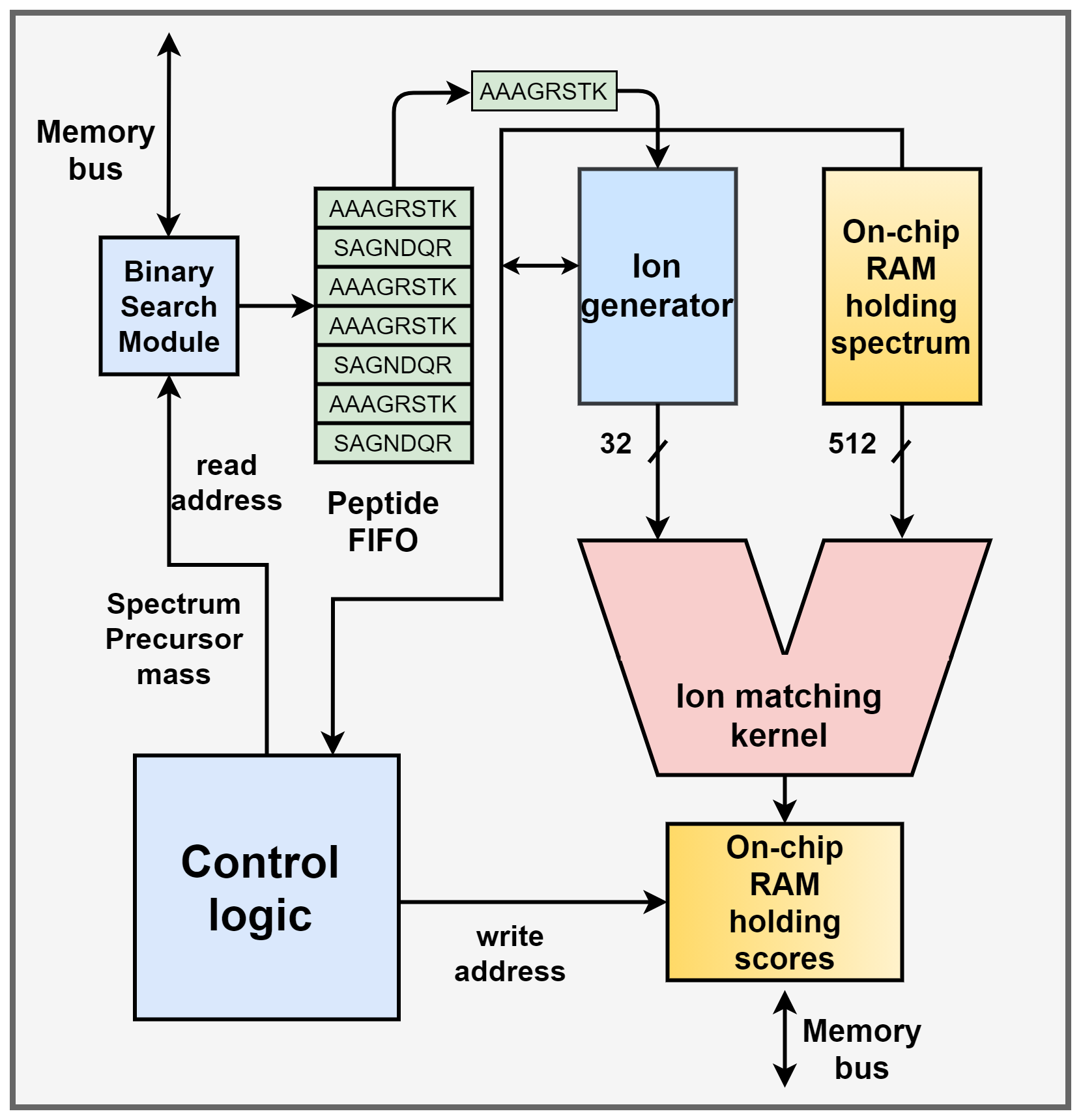}
\caption{Detailed internal construction of a single processing element. At the heart is the control logic which controls the function of all the sub-modules in the figure. Binary search module fetches a candidate peptide and stores it in a peptide FIFO. Ion generator reads the peptide and generates fragment ions. A 512 bit packet containing 16 32-bit ion mz and intensity pair values along with a 32-bit theoretical ion and intensity pair values are fed to the ion-matching kernel which finds the matching peak and stores the partial score in on-chip RAM.}
\label{fig:pe}
\end{center}
\end{figure}

Each PE takes over the computation of a single spectrum with all the candidate peptides. At the heart of a PE, sits a controller which determines the flow of computation as shown in Fig. \ref{fig:pe}. The controller copies an experimental spectrum in the form of m/z and ion intensity values from the external memory into on-chip RAM and starts the computation. Once the scores have been computed, they are collected in the on-chip RAM and a request for bus access is generated again to copy the scores into the DRAM.

\subsection{Bus Arbiter Design}
\begin{figure}[ht]
\begin{center}
\includegraphics[width=0.85\linewidth]{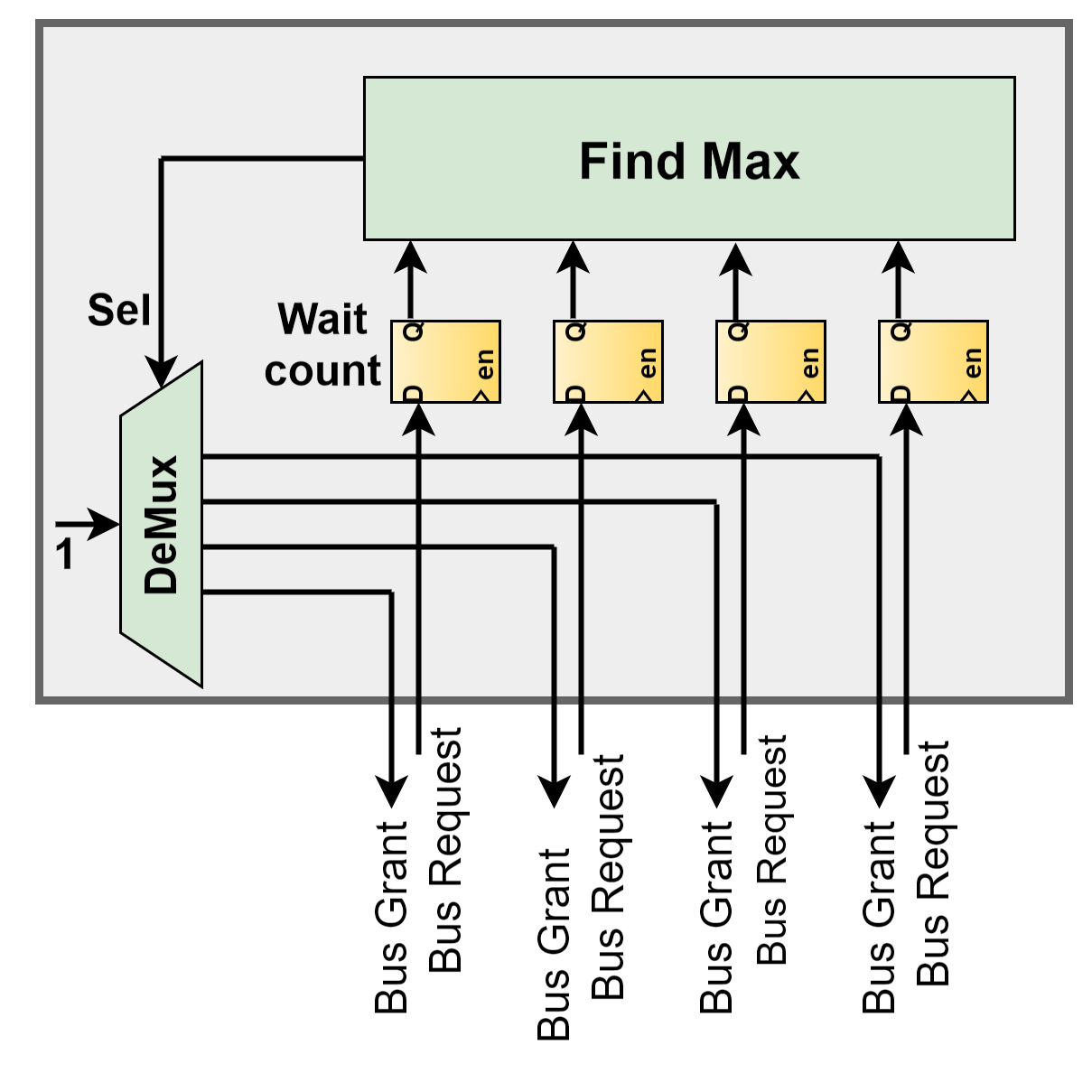}
\caption{Brief logic description of bus arbitration module. Bus request lines from all the PEs are coming into the arbiter. When a PE is denied service, its wait count register is incremented, dynamically increasing its priority for the next turn. Find max module is a comparator tree which finds which registers has the maximum value and grants access to the corresponding PE.}
\label{fig:arbiter}
\end{center}
\end{figure}

 To ensure load-balancing among the PEs, we designed a bus-arbiter shown in Fig. \ref{fig:arbiter} which aimed to minimize total wait time for all the PEs. All the PEs requiring access to the bus connect with the "bus request" signal which is connected to a wait counter register. The wait counter register keeps track of the wait time of every bus master so that the decision of contest for bus access is based on fairness i.e. access is granted to a master which has been wait for the longest. 

\subsection{Ion-Matching Kernel}

\begin{figure}[ht]
\begin{center}
\includegraphics[width=0.95\linewidth]{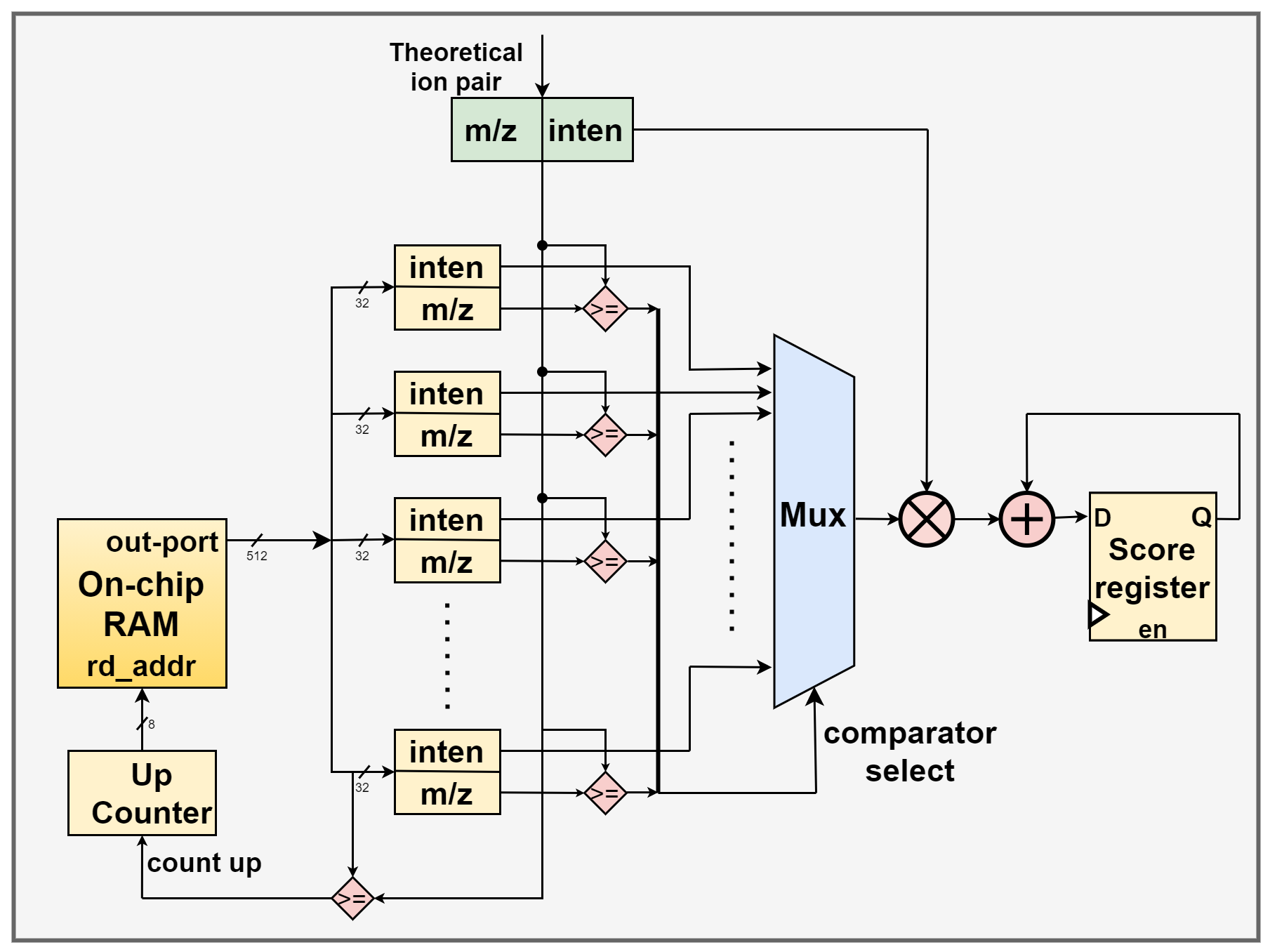}
\caption{The ion-matching circuit receives a 512 bit packet containing 16 experimental ions which are all compared with a theoretical ion in one cycle. The matched ions are multiplied and accumulated in the score register. If the theoretical ion is outside the range of current experimental ions, next packet is requested from the onchip-RAM by incrementing the counter. }
\label{fig:ion}
\end{center}
\end{figure}

To compute the dot product scores between experimental spectra and a candidate peptide, the processing element moves a 64-byte word from the on-chip RAM and a theoretically generated ion-pair from the candidate peptide to the peak-matching circuit. Each 64-byte word has 16 ion-pairs (using 16-bit floating point representation for intensity and 16 bit binary representation for m/z) from the experimental spectrum which is stored in 16 32-bit registers inside the peak-matching circuit. The m/z value of the theoretical ion is compared with all the experimental m/z values using a set of 16 parallel comparators as shown in Fig. \ref{fig:ion} and the corresponding matching intensity value of the peak is multiplied and accumulated in the output register. Once all the ions are traversed, the final Xcorr score is sent to the local on-chip RAM and the process repeats for the next candidate peptide. 

\section{Experiments}
\subsection{Methodology}
We designed the entire hardware using Intel Quartus Pro and Qsys system builder for Intel Stratix 10 FPGA. VHDL description was compiled using Quartus Pro to verify the maximum operable frequency of 200MHz. To evaluate the timing performance of our design, we implemented a cycle accurate simulator in python which mimicked the exact timing response of the hardware. In our simulator, we modeled each sub-module as a class whose data objects represented the internal and external signals of the module and a \textit{clock-event()} method which updated the signals whenever a clock edge occurred. \par
For our experiments, we used the PXD000612 dataset from PRIDE database which contained 90494 experimental spectra to score against human proteome dataset containing 669964 peptides. The experimental spectra were stored in the compressed sparse row (CSR) format with ion m/z value as the data index and ion intensity value as data element. Ion m/z values were stored in a 16 bit binary format and ion intensity values were represented using 16 bit half-precision floating point format. 

\begin{figure} 
    \centering
  \subfloat[\label{1a}]{%
       \includegraphics[width=0.45\linewidth]{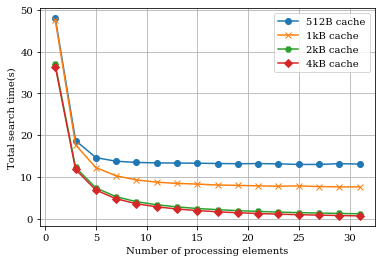}}
    \hfill
  \subfloat[\label{1b}]{%
        \includegraphics[width=0.45\linewidth]{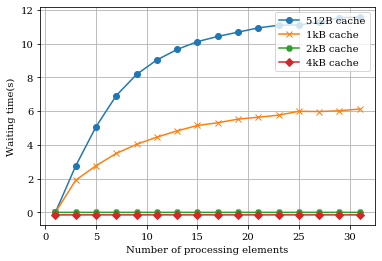}}
    \\
  \subfloat[\label{1c}]{%
        \includegraphics[width=0.45\linewidth]{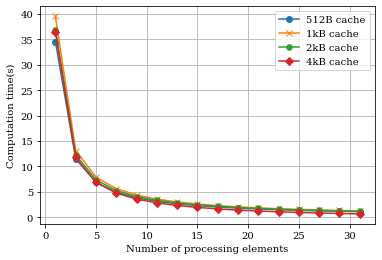}}
    \hfill
  \subfloat[\label{1d}]{%
        \includegraphics[width=0.45\linewidth]{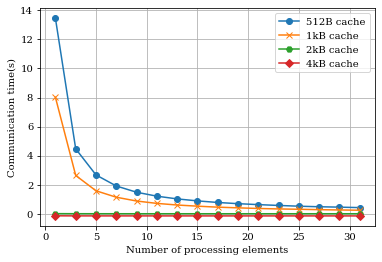}}
  \caption{(a), Total computation time vs number of instantiated PEs for cache size of 512B, 1kB, 2kB, and 4kB is shown. The search time decreases consistently with increasing number PEs for cache size 2kB and 4kB but its saturates for 512B and 1kB after 6 PEs due to increased memory requests. (b) Average synchronization time per PE in the system. This is the time spent by PE waiting for memory bus access. The wait time increases exponentially for cache size below 2kB. (c) Average computation time per PE in the system. This is the time spent on actual computation which decreases with the increasing number of PEs but is not affected by cache size. (d) Average I/O time per PE is also higher for cache size below 2kB. However there is little difference in performance between 2kB and 4kB.}
  \label{dse} 
\end{figure}

\subsection{Results}
The performance gains in our design come from a combination of optimizations which minimize DRAM accesses and allow input reuse by using an on-chip RAM as a local cache. To find the optimal cache size, we performed a design space exploration for four different cache sizes along with the number of instantiated PEs in the design. The results of these experiments are presented in Fig. \ref{dse}.
\par
We analyzed the performance of our design by elaborating the total processing time spent on computation and communication. To understand the effect of cache size, we further divide the communication time in terms of I/O and waiting time. We define these terms as \textbf{Average computation time}: average time each PE spends on computing dot product. \textbf{Average I/O time}: average time spent by each PE on DRAM read/write operations. \textbf{Average waiting time}: average time each PE spends on waiting to get access to system bus. The total processing time for dot product computations is shown in Fig. \ref{1a}, it is evident that increasing the number of PEs from 1 to 31 displayed significant speed up until 15 processing elements for a cache size of 2kB and 4kB, while for cache size below 2kB the speed up is plateaued after 6 processing elements. \par
 Fig. \ref{1b}, \ref{1c}, and \ref{1d} show the breakdown of the total processing time in terms of computation, I/O and waiting time for a single PE. Fig. \ref{1b} shows that waiting time is zero for 1 PE as memory bus is not being shared. As the number of PEs are increased, there is an almost exponential increase in the average waiting time of a PE for cache-sizes below 2kB. The waiting times for 2kB and 4kB cache stay constant even when 31 PEs are instantiated. The average computation time per PE is not impacted by the size of cache, but it decreases sharply when processing elements are increased to 11. Fig. \ref{1d} shows that the total I/O time i.e. total number of DRAM accesses are orders of magnitude greater for cache size below 2kB. Table \ref{tab0} further illustrates that increasing cache-size from 1kB to 2kB results in 600$\times$ reduction in the average I/O time.\par

\begin{table}[htbp]
\caption{Effect of cache size on average I/O and synchronization time while using 16 PEs}
\begin{center}
\begin{tabular}{|c|c|c|c|c|}
\hline

\textbf{Cache} & \textbf{I/O} & \textbf{Waiting} & \textbf{Total communication}\\
\textbf{size} & \textbf{time} & \textbf{time} & \textbf{time}\\
\hline
        512B & 1.01s & 11.57s & 12.58s\\
\hline
        1kB  & 0.52s  & 5.19s & 5.71s \\
\hline
        2kB & 0.86ms & 2.2ms & 3.06ms\\
\hline 
        4kB  & 0.84ms & 2.1ms & 2.95ms\\
        
\hline
\end{tabular}
\label{tab0}
\end{center}
\end{table}

Based on our experiments, in our final design we instantiated 16 processing elements to achieve maximum performance from the system. We compared the total search time of our design with Crux\cite{mcilwain2014crux} for 6 different values of precursor mass window. Table \ref{tab1} presents the total run-time of Crux running on a 3.6GHz Intel i7-4970 processor with 16GB of system memory and the run-time of our proposed hardware accelerator running at 200MHz clock frequency.

\begin{table}[htbp]
\caption{Run-time Comparison of FPGA accelerator with Crux running on 3.6GHz Intel i7-4970 using 8 threads and 16GB memory.}
\begin{center}
\begin{tabular}{|c|c|c|c|c|}
\hline
\textbf{Precursor mass}&                  \textbf{Dot product}        &   & \textbf{Hardware}    &  \textbf{Relative}\\
\textbf{Tolerance (Da)} & \textbf{operations} &   \textbf{\textit{Crux}} & \textbf{Accelerator} &  \textbf{ Speed-up} \\
\hline
        1.5 & 162.79M & 53.2s & 1.25s & 42 \\
\hline
        3  & 325.41M & 75s & 2.45s  & 30 \\
\hline
        5 & 541.42M & 86s & 4.10s & 21\\
\hline 
        10  & 1.07B & 139s & 7.782s & 20\\
\hline 
        25  & 1.99B & 304s & 20.75s & 15\\
\hline 
        50  & 3.076B & 648s & 39.6s & 16\\
        
\hline
\end{tabular}
\label{tab1}
\end{center}
\end{table}

\section{Conclusion}
In this paper we designed, and developed an efficient communication-avoiding micro-architecture. By using extensive experimentation, we demonstrated the applicability of custom hardware design approach to accelerate crucial memory bound problems in MS based omics. We presented optimizations for input reuse at all stages of the computation including cache implementation, pre-fetching, and input broadcasting. Although the system was designed for SEQUEST, it can easily be applied for other scoring techniques which involve dot product computation with little modification. Our simulation results suggest that our design is scalable for up-to 32 PEs with linear speed-ups. In future, we plan to extend this work to include the complete peptide identification process involving protein digestion, and build towards a general purpose proteomics processor.

\section*{Acknowledgment}
Research reported in this paper was supported by NIGMS of the National Institutes of Health under award number: R01GM134384. Fahad Saeed was further supported by the National Science Foundations (NSF) under the Award Numbers NSF CAREER OAC-1925960. The content is solely the responsibility of the authors and does not necessarily represent the official views of the National Institutes of Health or the National Science Foundation. 
\bibliographystyle{plain}
\bibliography{paper}
\end{document}